\documentclass[12pt]{article}
\usepackage{amsmath}
\usepackage{amsfonts}
\usepackage{amssymb}
\usepackage{srcltx}

\numberwithin{equation}{section}

\providecommand*{\iu}%
{\ensuremath{\mathrm{i}\,}}

\def\br{\begin{eqnarray}}
\def\er{\end{eqnarray}}
\def\be{\begin{equation}}
\def\ee{\end{equation}}
\def\nonu{\nonumber}
\def\lb{\lbrack}
\def\rb{\rbrack}
\def\({\left(}
\def\){\right)}

\relax

\def\a{\alpha}
\def\b{\beta}

\def\g{\gamma}

\def\l{\lambda}
\def\L{\Lambda}

\def\o{\over}
\def\O{\Omega}

\def\pa{\partial}

\def\rh{\rho}

\def\s{\sigma}
\def\t{\tau}

\def\lie{{\cal{G}}}

\def\PTP#1#2#3{{\sl Prog. Theor. Phys.} {\bf #1} (#2) #3}

\def\IJMPA#1#2#3{{\sl Int. J. Mod. Phys.} {\bf A#1} (#2) #3}

\def\JPA#1#2#3{{\sl J. Physics} {\bf A#1} (#2) #3}

\begin{document}
\noindent

\begin{center}
{\large\bf Construction of Type-II Backlund Transformation for the mKdV Hierarchy}
\end{center}
\normalsize
\vskip .4in

\begin{center}
J.F. Gomes, A.L. Retore and A.H. Zimerman 

\par \vskip .1in \noindent
Instituto de F\'{\i}sica Te\'{o}rica-UNESP\\
Rua Dr Bento Teobaldo Ferraz 271, Bloco II, \\
01140-070, S\~ao Paulo, Brazil\\
\par \vskip .3in

\end{center}
\baselineskip=.6cm
\begin{abstract}

From an algebraic construction of the mKdV hierarchy  we observe that  the space component of the Lax operator  play a role of an universal 
algebraic object. This fact induces the universality of a gauge transformation  
that relates two field configurations of a given member of the hierarchy.  Such gauge transformation generates the Backlund transformation (BT). 
In this paper we propose a systematic construction of     Backlund Transformation  for the entire mKdV hierarchy form the known Type-II BT of the sinh-Gordon theory.
 
We explicitly construct the BT  of  the first few integrable models associated to positive and negative grade-time evolutions.  
Solutions  of these transformations for several cases   describing  the transition from vacuum-vacuum and the vacuum to one-soliton solutions which  determines 
the  value for the auxiliary field and the 
the Backlund parameter respectively, independently of the model.
The same follows for   the scattering of two one-soliton solutions.  The resultant    delay is determined by a   condition independent of the model considered.

\end{abstract}

\section{Introduction}

 Backlund transformation (BT)  was introduced  a long time ago and  relates two solutions of
certain  non-linear equation of motion in a  lower order  differential equations.   Backlund transformation  appears as 
a peculiar feature of integrable models and generates a construction of an infinite sequence of  soliton solutions from
a non-linear superposition principle (see  \cite{rogers} for  a review).

More recently BT  has been employed to describe  integrable defects in the sense that  two
solutions of an integrable model  may be interpolated by a defect at certain spatial position.
The integrability is preserved if, at the defect position,  the two solutions  are described by a Backlund transformation.
The formulation is  such that,   canonical linear momentum and energy are  no longer conserved
and modifications to ensure its conservation need to be added  to take into account the  contribution of the defect \cite{cor-bow} .
A few  well known (relativistic) integrable models as the sine (sinh)-Gordon \cite{cor-zamb1}, Lund-Regge \cite{bow} 
and other  (non
relativistic)  models  as Non-Linear Schroedinger (NLS), mKdV, etc  have been studied
within such context \cite{cor-zamb2} .

The first type of Backlund transformation involves only the fields of the theory and is called type I.
 In particular, it may be observed that the {\it  space component}  of the  type I Backlund  transformation for the mKdV   and sinh-Gordon  
 equations  coincides for their corresponding fields \cite{rogers}.
 On the other hand, these two equations are known to belong to the same integrable  hierarchy which  is characterized by a common   (space) Lax operator.
 In fact, an infinite  number of integrable equations of motion  may be systematically constructed  from the same Lax operator and the  set of equations are 
 known as an integrable hierarchy (see for instance \cite{guil}).  Each equation of motion describes the time evolution of a field according to some integer graded 
 object of a Lie algebraic origin.
 In \cite{ghent} it was observed that all integrable equations  within the same  hierarchy share the same  space  component of the
 type I Backlund transformation.  They differ  by the time component Backlund transformation which was   constructed from  the equations of motion.

 More recently a new type of Backlund transformation involving  auxiliary fields was shown to be compatible with the equations of motion for 
 the sine (sinh)-Gordon and Tzitzeica models \cite{cor-zamb2}.
 These are known as type II Backlund transformation.  In \cite{thiago} it was shown that they may be constructed  from gauge transformation  relating two field
 configurations of the same equation of motion.

 The purpose of this paper is  to extend such results  to type II Backlund transformation to all integrable equations of the mKdV hierarchy. 
 Here we observe that 
  the  space 
 component of the type II Backlund transformation remains the same  {    within the hierarchy} and have developed a systematic construction for the time 
 component for several cases.
 
 In section 2 we review the construction of integrable hierarchies  from the algebraic approach.  
 We show the mKdV and sinh-Gordon  models represent the first  
examples of the positive ($t_N=t_3$) and negative ($t_N=t_{-1}$) sub-hierarchies respectively.  
Higher (Lower) grade  examples are also developed.  In section 3 we discuss the type II Backlund transformation for the sinh-Gordon (SG) models proposed in \cite{cor-zamb2} and 
present a gauge matrix $K$  that interpolates  between two   SG solutions \cite{thiago}.   Such gauge matrix  acts  on a two dimensional gauge potentials  
of the zero curvature representation.

In section 4, assuming $K$ to be {\it  universal within all models of the hierarchy},  we extend  the construction to the time component BT for the positive sub-hierarchy.  
We consider explicitly the mKdV ($t_N=t_3$)  and the next nontrivial equation for $t_N = t_5$.  We show that the compatibility of the two components of the BT 
indeed generate the correct equations of motion.  In section 5 we  develop the formalism for the negative grade  sub-hierarchy and   develop cases  for $t_N, N=-3,-5$. 
Again $K$ plays the role of universal object 
which leads to a compatible  pair of BT.

Finally, in section 6 we discuss the first three simplest solutions for the BT. The first solution consists of two vacuum solution which fixes  a condition for the auxiliar field $\L$.
The second describes the transition between the vacuum and one-soliton solutions which establishes   a condition for the Backlund parameter.We have considered several cases where $t_N, N=-1, \pm 3, \pm 5$  and in all of them, the very same condition arises.
The third  solution consists of the scattering of two one-soliton solutions where the delay (phase shift) $R$  is determined. 
Again   considering the   cases where $t_N = -1, \pm 3,\pm 5$ we found the very same solution for $R$.   These examples indicate  an universality of  Backlund 
parameters and delays within  the different models of the hierarchy.

\section{The mKdV Hierarchy}

In ref. \cite{guil} it is  explained how the mKdV hierarchy can be constructed from a basic {\it  universal object}   known as  Lax operator $L$ defined as
\br
L= \pa_x + E^{(1)} + A_0
\label{lax}
\er
where  the fundamental Lie algebraic  elements  $E^{(1)} =E_{\a}^{(0)} + E_{-\a}^{(1)}$ and $A_0 = v(x,t)h^{(0)}$ are  obtained from a  decomposition of an 
affine (centerless) $\hat {\lie} =\hat {sl}(2)$ Kac Moody  algebra
\br
\lb h^{(m)}, E_{\pm \a}^{(n)} \rb = \pm 2 E_{\pm \a}^{(m+n)}, \qquad \lb E_{ \a}^{(m)}, E_{- \a}^{(n)}\rb = h^{(m+n)}\qquad \lb d, T_i^{(n)} \rb = nT_i^{(n)}  
\label{1}
\er 
into graded subspaces, $\hat {\lie} = \oplus_a \lie_a, a \in Z$.  
Here  $ T_i^{(n)}$ denotes either $   h^{(n)} $ or $ E_{\pm \a}^{(n)} $.
For the mKdV hierarchy, such decomposition is  induced by the grading
operator (principal gradation) $Q= 2 d + 1/2 H^{(0)}$ such that the even and odd subspaces are  respectively given by
\br 
\lie_{2n} = \{ h^{(n)}\}\qquad 
\lie_{2n+1} = \{ E_{a}^{(n)} + E_{-\a}^{(n+1)},   E_{a}^{(n)} - E_{-\a}^{(n+1)}  \}  
\er
i.e.,
\br
\lb Q, \;  \lie_{2n+1}\rb = (2n+1) \lie_{2n+1}, \qquad \lb Q, \; \lie_{2n}\rb = 2n \lie_{2n}.
\er
and  $\lb \lie_a, \lie_b \rb \in \lie_{a+b}$. Furthermore,   the choice of the constant grade one element   $E^{(1)}$,  defines the Kernel, i.e.,  
${\cal K} = \{ \forall x \in \hat {\lie}, \quad  [ x,   E^{(1)}]=0 \} $ and induces a second decomposition of the affine algebra $\hat \lie ={\cal K} \oplus {\cal M} $.
  It follows that for the mKdV hierarchy the Kernel has grade $2n+1$ and is generated by the combination
  \br
  {\cal K}= {\cal K}_{2n+1} = \{  E_{a}^{(n)} + E_{-\a}^{(n+1)}\}.
  \er
 ${\cal M} $ is its complement and satisfies
\br
\lb  {\cal K} , {\cal K} \rb \subset  {\cal K}, \qquad
\lb  {\cal K} , {\cal M} \rb \subset  {\cal M}, \qquad
\lb  {\cal M} , {\cal M} \rb \subset  {\cal K}.
\nonu
\er

In general the Lax operator (\ref{lax}) is constructed sistematically   from $A_0 \in {\cal M}_{0} = {\lie_0} \cap {\cal M}$ and is parametrized  by the physical fields of the theory.

Equations of motion for the {\it positive mKdV sub-hierarchy} 
are given by the zero curvature representation
\br
\lb \pa_x + E^{(1)} + A_0,\quad  \pa_{t_N} + D^{(N)} + D^{(N-1)} + 
\cdots +D^{(0)} \rb = 0. \label{5}
\er
The  solution of eq. ( \ref{5}) may be  systematically constructed  by considering $D^{(j)} \in \lie_j$
and   decomposed according to the graded structure as 
\br
\lb E^{(1)}, D^{(N)}\rb &=0& \label{6.1} \\
\lb E^{(1)}, D^{(N-1)}\rb + \lb A_0, D^{(N)}\rb +\pa_x D^{(N)} &=0& \nonu \\
\vdots &=& \vdots \nonu \\
\lb A_0, D^{(0)}\rb + \pa_x D^{(0)} - \pa_{t_{N}}A_0 &=&0. 
\label{6}
\er
The unknown $D^{(j)}$'s can be solved starting from the highest to the lowest grade projections as functionals of $A_0 $ and its $x-$ derivatives.  
Notice that, in particular   the highest grade equation, namely $\lb E^{(1)}, D^{(N)}\rb =0$  implies $D^{(N)} \in {\cal K}$ and henceforth $N=2n+1$. 
If we  consider the  fields of the theory to parametrize $ A_0 = v(x, t_N) h^{(0)}  \in {\cal M}_0$, 
the  equations of motion are obtained from the zero grade component  (\ref{6}). 
Examples are, 
\br
N=3 \qquad 4\pa_{t_3} v&=& \pa_x \(\pa_x^2v -2 v^3 \)\qquad  mKdV  \nonu \\
\label{t3} \\
\nonu \\
N=5 \qquad 16\pa_{t_5}v  &=& \pa_x \(   \pa_x^4v -10v^2(\pa_x^2 v) -10v(\pa_x v)^2+6v^5 \), \nonu \\
\label{t5} \\
\nonu \\
N=7 \qquad 64\pa_{t_7}v &=& \pa_x \(  \partial_{x}^6v-70 (\partial_{x}v)^{2}(\partial_{x}^2v)-42v(\partial_{x}^2v)^{2}-56 v(\partial_{x}v)(\partial_{x}^3v)\)\nonu \\  
&-&\pa_x \(14 v^{2}\pa_x^4v -140 v^{3}(\partial_{x}v)^{2}-70 v^{4}(\partial_{x}^2v)+20v^{7}  \)\nonu \\ 
\label{t7} \\
  \cdots etc \nonu 
\er 
For the {\it negative mKdV sub-hierarchy} let us propose the following form for the zero curvature representation 
\br
\lb \pa_x + E^{(1)} + A_0, \pa_{t_{-N}} + D^{(-N)} + D^{(-N+1)} 
+\cdots +D^{(-1)} \rb = 0.
\label{7}
\er
Differently  from the positive sub-hierarchy case, the lowest grade  projection now yields,
\br
\pa_x D^{(-N)} + \lb A_0, D^{(-N)}\rb = 0,
\nonu
\er
 a nonlocal equation for $D^{(-N)}$.   
 The proccess follows recursively until we reach the zero grade projection 
\br
\pa_{t_{-N}} A_0  - \lb  E^{(1)},D^{(-1)}\rb  = 0
\label{8}
\er
which yields the evolution equation for field $A_0$ according to time $t=t_{-N}$
  The simplest example is to take $N=1$ when the zero curvature decomposes into 
 \br
\pa_x D^{(-1)} + \lb A_0, D^{(-1)}\rb & = & 0, \nonu \\
\pa_{t_{-1}} A_0  - \lb  E^{(1)},D^{(-1)}\rb  & = & 0.
\label{9}
\er
In order to solve the first equation, we define the zero grade group element $B = \exp \( \lie_0 \) $ and  define
 \br
D^{(-1)} = B E^{(-1)} B^{-1}, \qquad A_0 = -\pa_x B B^{-1}, 
\label{10}
\er
where $E^{(-1)} = E_{\a}^{(-1)} + E_{-\a}^{(0)}$. Under such parametrization  the  first eqn. (\ref{9}) is automatically satisfied whilst  the second becomes the well known (relativistic ) Leznov-Saveliev equation,
\br
\pa_{t_{-1}} \(\pa_x B B^{-1}\) + \lb  E^{(1)}, B E^{(-1)}B^{-1} \rb =0
\label{11}
\er
which for $\hat{sl}(2)$ with principal gradation 
$Q= 2\l {{d}\o {d\l}} + {1\o 2} h$, yields the sinh-Gordon equation
\br
\pa_{t_{-1}} \pa_x \phi  = e^{2\phi} - e^{-2\phi}, \qquad  B= e^{-\phi h}.
\label{sg}
\er
where $t_{-1} = z, x = \bar z$ are the light cone coordinates and $  A_0 = -\pa_x B B^{-1} = v h \equiv  \pa_x \phi h$. 

 For higher values of $N=3,5,\cdots$ {\footnote{ For the negative sub-hierarchy  there is no restriction upon the values of $N$ to be odd.  In particular  integrable models 
 for $N$ even  present  non trivial vacuum solution as it was explicitly discussed  in ref. \cite{guil}}} we find
 \br
 \pa_{t_{-3}}\pa_x \phi &=& 4e^{-2\phi} d^{-1}\( e^{2\phi}d^{-1} (\sinh 2\phi)\) +4 e^{2\phi} d^{-1}\( e^{-2\phi}d^{-1} (\sinh 2\phi)\)
 \label{tm3}
 \er
 \br
 \pa_{{t_{-5}}}\pa_x \phi &=& 8 e^{-2\phi} d^{-1} \(e^{2\phi} d^{-1}\( e^{-2\phi} d^{-1}\( e^{2\phi}d^{-1} (\sinh 2\phi)\) + e^{2\phi} d^{-1}\( e^{-2\phi}d^{-1} (\sinh 2\phi)\)\)\) \nonu \\
 &+& 8 e^{2\phi} d^{-1} \(e^{-2\phi} d^{-1}\(e^{-2\phi} d^{-1}\( e^{2\phi}d^{-1} (\sinh 2\phi)\) + e^{2\phi} d^{-1}\( e^{-2\phi}d^{-1} (\sinh 2\phi)\)\)\), \nonu 
 \\
 \label{tm5}
 \er
 where $d^{-1}f = \int^x f(y)dy$.

\section{On the Sinh-Gordon Type II Backlund Transformation }

The Lax  pair $L= \pa_x + A_x$ in  (\ref{lax}) for the sinh-Gordon model where $v(x, t) = \pa_x \phi(x, t)  $ is specified by
\br
 A_{x} = \pa_{x} \phi h^{(0)} + E_{\a}^{(0)} + E_{-\a}^{(1)} = \begin{bmatrix} \pa_{x}\phi &1  \\ \l & -\pa_{x}\phi \end{bmatrix}
\label{laxsg}
\er
and for $t=t_{-1}$, we find  from (\ref{10}) 
\br
A_{t_{-1}} = B E^{(-1)} B^{-1} = e^{-2\phi}  E_{\a}^{(-1)} +  e^{2\phi}  E_{-\a}^{(0)} =  \begin{bmatrix} 0 &  {{1}\over {\l}}e^{-2\phi} \\   {e^{2\phi}} &0 \end{bmatrix}.
\label{xxx}
\er
The zero curvature  condition 
\br
\left[ \pa_t+ A_t , \pa_x + A_x \right] = 0
\label{zcc}
\er
for $t= t_{-1}$ leads to the sinh-Gordon eqn., 
\br
\pa_{t_{-1}} \pa_x \phi  - 2 \sinh (2\phi ) =0 \label{sg}
\er
in the light cone coordinates $(x, t_{-1})$.
In order to determine the type-II
 Backlund transformation for the sinh-Gordon eqn. (\ref{sg}) proposed in \cite{cor-zamb1}    a gauge matrix $K(\phi_1, \phi_2)$  such that 
 \br
 \pa_{\mu} K =K A_{\mu}(\phi_1) - A_{\mu}(\phi_2) K, \qquad x_{\mu} = t_{-1}, x
 \label{k}
 \er was constructed  in \cite{thiago} (see appendix B).  Let K be given by
 \br
 K = \begin{bmatrix} 1- {{e^q}\over {\l \s^2}} & {{e^{\L -p}\over {2\l \s}} }(e^q + e^{-q} +\eta)  \\ -{{2}\over {\s}}e^{p-\L} & 1- {{e^{-q}}\over {\l \s^2}} \end{bmatrix}
 \label{kmatrix}
 \er
 where 
 \br
 p=\phi_1 + \phi_2, \qquad q = \phi_1- \phi_2, \label{pq}
 \er
  $ \L (x, t_N)$ is an auxiliary field and $\s $  the Backlund parameter.
 Eqn. (\ref{k}) with $A_{\mu} = A_x$ and $K$ given by (\ref{laxsg}) and (\ref{kmatrix}) respectively leads to 
 \br
 \pa_x q &=& -{{1}\over {2\s}} e^{\L-p} (e^q +e^{-q} + \eta) -{{2}\over {\s}} e^{p-\L}, \label{type-iixa}\\
 \pa_x \L&=& {{1}\over {2\s}} e^{\L-p} (e^q -e^{-q} ).
 \label{type-iixb}
 \er
 For $\mu = t_{-1}$ in (\ref{k}) we find for $A_{t_{-1}}$ given by (\ref{xxx}),
 \br
 \pa_{t_{-1}} q &=& -{2\s} e^{-\L} - {{\s}\over {2}}e^{\L} (e^q +e^{-q} + \eta), \label{type-iita}\\
 \pa_{t_{-1}} (p- \L)&=& -{{\s}\over {2}} e^{\L} (e^q -e^{-q} ).
 \label{type-iitb}
 \er
 Eqns. (\ref{type-iixa}) -(\ref{type-iitb}) are the type II Backlund transformation for the sinh-Gordon introduced in \cite{cor-zamb1}
 in the sense that if we act with $\pa_{t_{-1}}$ in eqn. (\ref{type-iixa}) 
 and  (\ref{type-iixb}) we obtain,  using (\ref{type-iita}) and (\ref{type-iitb}) relations consistent with the sinh-Gordon equation for $\phi_1$ and $\phi_2$.  
 Conversely, acting with $\pa_{x}$ in eqn. (\ref{type-iita}) and  (\ref{type-iitb}) and   using (\ref{type-iixa}) and (\ref{type-iixb}) we obtain similar result.

 In particular, from eqns. (\ref{type-iixa}) and (\ref{type-iita})  we find the following expression for $\L$,
 \br
 e^{-\L} &=& -{{1}\over {2\s }} {{\s^2 \pa_x q -e^{-p} \pa_{t_{-1}} q }\over {e^{p} - e^{-p}}}, \qquad 
 e^{\L} = {{2}\over {\s }} {{\s^2 \pa_x q -e^{p} \pa_{t_{-1}} q }\over {(e^{p} - e^{-p}})(e^{q} + e^{-q} + \eta) }. \label{l1}
 \er
 \section{Type II Backlund Transformation for  the positive mKdV Sub-Hierarchy } 
\begin{itemize}
\item $t_N = t_3$

Based upon the fact that  the Lax operator $L$ in  (\ref{lax})  is common to the entire hierarchy  we propose the {\it conjecture} that the gauge matrix $K$  is also 
common  and provide the type II Backlund transformation to all members of the hierarchy. In particular this fact implies that the  
space component of the Backlund transformation is  the same for all members of the hierarchy.
In order to  support our conjecture we now consider the mKdV equation 
\br
4\pa_{t_3}v = \pa_x^3 v -6 v^2 \pa_x v, \qquad v = \pa_x\phi.  
\label{mkdv} 
\er
We shall consider $A_x $ given by (\ref{lax}) and (\ref{laxsg}) together with   $A_t = A_{t_3}$,
\br
A_{t_3} &=& D^{(3)} + D^{(2)} + D^{(1)} + D^{(0)} , \nonu \\ 
&=& E_{\a}^{(1)} + E_{-\a}^{(2)} + v h^{(1)} + {{1}\over {2}}(\pa_x v - v^2) E_{\a}^{(0)} 
- {{1}\over {2}}(\pa_x v + v^2)E_{-\a}^{(1)}\nonu \\
&+& {{1}\over {4}}(\pa_x^2v - 2 v^3)h^{(0)}\nonu \\
&=&  \begin{bmatrix} \l v +{{1}\over {4}} \pa_x^2v -{{1}\over {2}}v^3 & \l -{{1}\over {2}}v^2 +{{1}\over {2}}\pa_x v   \\ \l^2 -{{\l}\over {2}}v^2 -{{\l}\over {2}}\pa_xv & -\l v -{{1}\over {4}} \pa_x^2v +{{1}\over {2}} v^3 \end{bmatrix}
\label{a3}
\er
to generate the mKdV equation (\ref{mkdv}) from the zero curvature representation (\ref{zcc}).
 
 In order to derive the time component of the type II Backlund transformation  for the mKdV equation  we employ  eqn. (\ref{k})  with $x_\mu = t_3$, i.e.,
 \br
 \pa_{t_3} K =K A_{t_3}(\phi_1) - A_{t_3}(\phi_2) K
 \label{k3}
 \er
 where the gauge matrix $K$ is given by (\ref{kmatrix}).   
 
 The matrix element 11 of the above eqn. leads to:
  \br
  \l^{-1} &:& 4\pa_{t_3}(\phi_1 - \phi_2) = \pa_x^3 (\phi_1 - \phi_2) -2 ((\pa_x \phi_1)^3 - (\pa_x \phi_2)^3 ) \label{1.4}\\
   \l^{0} &:& \s^2 \pa_x^2(v_2 -v_1) -2\s^2 (v_2^3 - v_1^3) =4(v_2-v_1)e^{q} - 4\s (v_2^2 - \pa_x v_2)e^{p-\L}\nonu \\ 
   &-&\s (v_1^2 +\pa_x v_1)(e^q +e^{-q} +\eta)e^{\L -p} \label{1.5}\\
   \l^{1} &:& \pa_x q = -{{1}\over {2\s}} e^{\L-p}(e^q+e^{-q} + \eta)- {{2}\over {\s}} e^{p-\L}
   \label{1.6}
 \er 
 Eqn. (\ref{1.4}) is trivially satisfied  for $v_1 = \pa_x\phi_1$ and $v_2 = \pa_x\phi_2$ satisfying the mkdV equation (\ref{mkdv}).  
 Eqn. (\ref{1.6}) coincides with eqn. (\ref{type-iixa}) and represents the  space component of the Backlund transformation for the mKdV. 
 Acting twice with $\pa_x$ in (\ref{1.6}) and using (\ref{type-iixb}), after some tedious calculation it can be shown that (\ref{1.5}) is  identically satisfied.

 The matrix   elements 12 and 21  of (\ref{k3}) yields respectively  
 \br
  \l^{-1} &:& \s \pa_{t_3}\L (e^q +e^{-q} +\eta) +\s \pa_{t_3} q(e^q-e^{-q}) \label{1.7}\nonu \\
  &=& \((v_1^2 - \pa_x v_1)e^{p+q} - (v_2^2 - \pa_x v_2)e^{p-q}\)e^{-\L} \label{1.8}\\
   \l^{0} &:& \s^2 \pa_x (v_1-v_2)- \s^2 (v_1^2-v_2^2) = 2(e^q -e^{-q}) \nonu \\
   &+& \s (v_1+v_2) (e^q+e^{-q} +\eta)e^{\L-p} \\
   \l^{1} &:&  0=0
   \label{1.9}
 \er 
 and
 \br
  \l^{0} &:&  4\s \pa_{t_3} \L = (v_1^2+\pa_xv_1)e^{\L-q-p} - (v_2^2 + \pa_xv_2)e^{\L+q-p} \label{1.10}\\
   \l^{1} &:&\s^2 \pa_x (v_1-v_2) + \s^2 (v_1^2 -v_2^2) = 2 (e^q -e^{-q}) -4\s (v_1+v_2)e^{p-\L}\label{1.11} \\
   \l^{2} &:& 0=0
   \label{1.12}
 \er
 Eqns. (\ref{1.8}) and (\ref{1.11})  are direct consequences of eqns. (\ref{type-iixa}) and (\ref{type-iixb}). 
 The matrix  element 22 now yields,
 \br
 \l^{-1} &:& 4\pa_{t_3}q = \pa_x^2 (v_1-v_2) - 2 (v_1^3 -v_2^3), \label{1.13}\\
 \l^{0}  &:& \s^2 \pa_x^2(v_1-v_2) -2\s^2 (v_1^3-v_2^3) =4\s (v_1^2-\pa_xv_1)e^{p-\L} \nonu \\
 &+& \s (v_2^2+\pa_x v_2)(e^q+e^{-q}+\eta)e^{-p+\L} + 4 (v_1 -v_2)e^{-q}  \label{1.14}
 \er
Eqn. (\ref{1.13}) is the mKdV equation for  $v_1 = \pa_x\phi_1$ and $v_2 = \pa_x\phi_2$.  
Acting  with $\pa_x^2$ in eqn. (\ref{type-iixa}) and employing (\ref{type-iixb}) it can be shown that (\ref{1.14}) is trivially satisfied.
 It then follows   from (\ref{1.10}) in (\ref{1.7}) that the following pair of eqns.,
 
 \br
 16\s^{3}\pa_{t_{3}}q &=& e^{\L-p}\left(e^{q}+e^{-q}+\eta\right)\left[2 \s^{2} (\pa^{2}_{x}p+\pa^{2}_{x}q)+\s^{2} \left(\pa_{x}p+\pa_{x}q\right)^{2}-8e^{q}\right]+\nonu \\
&+& 4 e^{p-\L}\left[-2 \s^{2} (\pa^{2}_{x}p+\pa^{2}_{x}q)+\s^{2} \left(\pa_{x}p+\pa_{x}q\right)^{2}-8e^{-q}\right]+\nonu \\
&+& 16 \s \pa_{x}p\left(e^{q}+e^{-q}+\eta\right)
\label{1.15}
\er
\br
 4\s \pa_{t_3} \L &=& (v_1^2+\pa_x v_1)e^{\L-q-p} - (v_2^2 + \pa_x v_2)e^{\L+q-p} \label{1.16}
 \er
 together with (\ref{type-iixa}) and (\ref{type-iixb})
correspond to   the type-II Backlund transformation for the mKdV equation.  
Acting with $\pa_x $ in eqn. (\ref{1.15})  and $\pa_{t_3}$ in (\ref{type-iixa}) we find in both cases consistency with the mKdV equation (\ref{mkdv}).

Further, from (\ref{type-iixa}) and (\ref{1.15}) we find
\br
{{4}\over {\s}}e^{-\L} &=&- {{2\s^2 \pa_{{t_3}}q + \pa_xq \( \s^2 (\pa_x \phi_1)^2 + 2 (e^q + \eta ) + \s^2 \pa_x^2 \phi_1\) - 4 \pa_x \phi_1 ( e^q + e^{-q} + \eta )}
\over {e^p (\s^2\pa_x^2 \phi_1 - (e^q - e^{-q}))}}, \nonu \\
\label{l1a}\\
{{4}\over {\s}}e^{\L} &=&{{ {8 \s^2 \pa_{t_3}}q +\pa_x q\(4 \s^2(\pa_x \phi_1)^2 + 8 (e^q + \eta) - 4 \s^2 \pa_x^2 \phi_1 \)  - 16 \pa_x \phi_1 (e^q + e^{-q} + \eta )}\over
{ e^{-p} ( e^q + e^{-q} + \eta )( \s^2 \pa_{x}^2 \phi_1 - ( e^q - e^{-q} ))}} \nonu \\
\label{l1b}
\er
 
 \item $t_N=t_5$
 
 For $t=t_5$ we find
 \br 
16 \s^{5}\pa_{t_{5}}q =4Ae^{p-\L}+B(e^{q}+e^{-q}+\eta)e^{\L-p} +C
\label{back5}
\er
\br 
16\s \pa_{t_{5}}\L &=& \(\pa^{4}_{x}\phi_{1}-(\pa^{2}_{x}\phi_{1})^{2}+2(\pa_{x}\phi_{1})(\pa^{3}_{x}\phi_{1})-6(\pa_{x}\phi_{1})^{2}(\pa^{2}_{x}\phi_{1})-
3(\pa_{x}\phi_{1})^{4}\)e^{\L-q-p}\nonu \\
                  & -& \(\pa^{4}_{x}\phi_{2}-(\pa^{2}_{x}\phi_{2})^{2}+2(\pa_{x}\phi_{2})(\pa^{3}_{x}\phi_{2})-6(\pa_{x}\phi_{2})^{2}(\pa^{2}_{x}\phi_{2})-
                  3(\pa_{x}\phi_{2})^{4}\)e^{\L+q-p}\nonu \\
                \label{back5lambda}  
\er
with
\br 
A &=& 8(1-\eta e^{q}-\eta^{2})+4\s^{2}(e^{q}+\eta)(\pa^{2}_{x}\phi_{1}-(\pa_{x}\phi_{1})^{2})+\nonu\\
    & &-\s^{4}\left[(\pa^{2}_{x}\phi_{1})^{2}+\pa^{4}_{x}\phi_{1}-2(\pa_{x}\phi_{1})(\pa^{3}_{x}\phi_{1})-6(\pa_{x}\phi_{1})^{2}(\pa^{2}_{x}\phi_{1})+3(\pa_{x}\phi_{1})^{4} \right], \nonu \\
\er
\br 
B &=&  8(1-\eta e^{-q}-\eta^{2})-4\s^{2}(e^{-q}+\eta)(\pa^{2}_{x}\phi_{1}+(\pa_{x}\phi_{1})^{2})+\nonu\\
    & &+\s^{4}\left[-(\pa^{2}_{x}\phi_{1})^{2}+\pa^{4}_{x}\phi_{1}+2(\pa_{x}\phi_{1})(\pa^{3}_{x}\phi_{1})-6(\pa_{x}\phi_{1})^{2}(\pa^{2}_{x}\phi_{1})-3(\pa_{x}\phi_{1})^{4} \right], \nonu \\
\er
and
\br 
C=-32\s\eta(\pa_{x}\phi_{1})(e^{q}+e^{-q}+\eta)+8\s^{3}(\pa^{3}_{x}\phi_{1}-2(\pa_{x}\phi_{1})^{3})(e^{q}+e^{-q}+\eta).
\er
If we act with $\pa_{x}$ in the equation \eqref{back5} and use (\ref{type-iixa}) and (\ref{type-iixb})  we recover the equation of motion \eqref{t7}.
Conversely, acting  $\pa_{t_{5}}$ in the equation 
(\ref{type-iixa}) and using  \eqref{back5} \eqref{back5lambda} we obtain the same result, \eqref{t7}.

Using  \eqref{back5} together with (\ref{type-iixa}) we find the following expression for $\L$.
\br 
e^{\L}=\frac{16\s^{5}\pa_{t_{5}}q+2\s A\pa_{x}q-C}{e^{-p}(B-A)(e^{q}+e^{-q}+\eta)}, \qquad 
 e^{-\L}=\frac{16\s^{5}\pa_{t_{5}}q+2\s B\pa_{x}q-C}{4e^{p}(A-B)}
\er

 \end{itemize}
 
 \section{Type-II BT for  Negative Grade mKdV Sub-Hierarchy}
 \begin{itemize}
 \item $t_N=t_{-3}$
 
 We now derive the Type-II Backlund transformation for the $t=t_{-3}$ equation, 
 \br
 \pa_{t_{-3}}\pa_x \phi &=& 4e^{-2\phi} d^{-1}\( e^{2\phi}d^{-1} (\sinh 2\phi)\) +4 e^{2\phi} d^{-1}\( e^{-2\phi}d^{-1} (\sinh 2\phi)\)
 \label{tm3}
 \er
 obtained from the zero curvature representation (\ref{zcc}) with $A_x$ given by (\ref{laxsg})  and 
 \br
 A_{t_{-3}}(\phi) &=& e^{-2\phi}E_{\a}^{(-2)} + e^{2\phi}E_{-\a}^{(-1)} - 2 I(\phi) h^{(-1)} \nonu \\
 &-& 4 e^{-2\phi} \int^x e^{2\phi} I(\phi) E_{\a}^{(-1)} +4 e^{2\phi} \int^x e^{-2\phi} I(\phi) E_{-\a}^{(0)}
   \er
   where 
   \br
   I_i(\phi_i)= \int^x \sinh (2\phi_i (y)) dy .\label{i}
 \er
 The Backlund is generated by 
 \br
 \pa_{t_{-3}} K =K A_{t_{-3}}(\phi_1) - A_{t_{-3}}(\phi_2) K
 \label{k3}
 \er
 where the gauge matrix $K$ is the same given by (\ref{kmatrix}).  It therefore follows 
 that the matrix element 11 of the above eqn. leads to:
  \br
  \l^{-2} &:& I_1-I_2 = -{{\s e^{\L}}\over {4}} (e^q +e^{-q} +\eta ) - \s e^{-\L}    \label{11-2a}\\
  \l^{-1} &:& \pa_{t_{-3}} q =2\s^2 (I_1-I_2)e^{-q} -2 \s e^{\L}(e^q +e^{-q}+\eta )\int^x I_1 e^{-2\phi_1} \nonu \\
  &+& 8 \s  e^{-\L}\int^x I_2e^{2\phi_2} 
   \label{11-2b}
 \er 
 the matrix element 22   to 
 \br
  \l^{-2} &:&  \pa_{t_{-3}}q = 2\s^2 (I_1-I_2) e^{q} -2\s e^{\L} ( e^q +e^{-q} +\eta )\int^x I_2 e^{-2\phi_2} \nonu \\
  &+& 8\s e^{-\L} \int^x I_1e^{2\phi_1}  
   \label{22}
 \er 
  The  element 12 leads to
 \br
  \l^{-3} &:&   (I_1+I_2) (e^q +e^{-q} +\eta) = \s e^{-\L} (e^q -e^{-q})\nonu \\
  &-&{{4}\over {\s}} e^{-\L} \int^x (I_1 e^{2\phi_1} -I_2e^{2\phi_2}) 
  \label{12-2a}
  \er
  \br
  \l^{-1} &:&  \pa_{t_{-3}}(\L-p)(e^q+e^{-q}+ \eta)  + \pa_{t_{-3}}q (e^q -e^{-q})\nonu \\
  &=&-8\s e^{-\L} ( e^{-q} \int^x  I_1 e^{2\phi_1} -e^{q} \int I_2e^{2\phi_2} )   
   \label{12-2b}
 \er 
 Element 21 yields,
 \br
  \l^{-1} &:&  I_1+I_2 = -{{\s}\over {4}}e^{\L} (e^q-e^{-q}) + {{e^{\L}}\over{\s}} \int^x ( I_1 e^{-2\phi_1} - I_2e^{-2 \phi_2} ) \label{21-2a}\\
  \l^{-1} &:& \pa_{t_{-3}} (p-\L) = -2\s e^{\L+q}\int^x I_1e^{-2\phi_1} +2\s e^{\L-q}\int^x I_2 e^{-2\phi_2}     \nonu \\
   \label{21-2b}
 \er

 Acting with $\pa_x$ in eqns. (\ref{11-2a}),( \ref{12-2a}) and (\ref{21-2a}) and using (\ref{type-iixa}) and (\ref{type-iixb}) with  we find that they are identically satisfied.
 It therefore follows that the type-II Backlund transformation  for the eqn. of motion (\ref{tm3}) is given by
 \br 
 \pa_{t_{-3}} (p-\L) &=& -2\s e^{\L+q}\int^x I_1e^{-2\phi_1} +2\s e^{\L-q}\int^x I_2 e^{-2\phi_2}  \label{tm3a}
 \er
 \br
\pa_{t_{-3}} q &=&2\s^2 (I_1-I_2)e^{-q} -2 \s e^{\L}(e^q +e^{-q}+\eta )\int^x I_1 e^{-2\phi_1} + 8 \s  e^{-\L}\int^x I_2e^{2\phi_2}, \nonu \\
   \label{tm3b}
 \er
 \br 
  \pa_{t_{-3}}q = 2\s^2 (I_1-I_2) e^{q} &-&2\s e^{\L} ( e^q +e^{-q} +\eta )\int^x I_2 e^{-2\phi_2} +8\s e^{-\L} \int^x I_1e^{2\phi_1} \nonu\\
  \label{tm3c}
  \er
  \br
  \pa_{t_{-3}}(\L-p)(e^q+e^{-q}+ \eta) & +& \pa_{t_{-3}}q (e^q -e^{-q})\nonu \\
  &=&-8\s e^{-\L} ( e^{-q} \int^x  I_1 e^{2\phi_1} -e^{q} \int I_2e^{2\phi_2} ) \nonu \\
  \label{tm3d} 
 \er
 together with 
 \br
 \pa_x \L&=& {{1}\over {2\s}} e^{\L-p} (e^q -e^{-q} ), \label{txa}\\
 \pa_x q &=& -{{1}\over {2\s}} e^{\L-p} (e^q +e^{-q} + \eta) -{{2}\over {\s}} e^{p-\L}, \nonu \\
 \label{txb}
 \er
 Equations (\ref{tm3a}),  (\ref{tm3b}) and (\ref{txa}), (\ref{txb}) are compatible in the sense that  acting with $\pa_{t_{-3}}$ in  (\ref{txa}),  (\ref{txb}) we recover 
 the same equation  of motion (\ref{tm3}). Conversely, applying $\pa_{x}$ on equations \eqref{tm3b}  and  \eqref{tm3c} we obtain the equation of motion \eqref{tm3}
  after adding both results and using the equations \eqref{txa} and \eqref{txb}. 
 
From the above eqns. we find 
\begin{equation}
e^{\Lambda}=-\frac{1}{2\s}\frac{(\pa_{t_{-3}}q) e^{p}+4\s^{2}(\pa_{x}q)\int^{x}I(\phi_{2})e^{2\phi_{2}}dy-2\s^{2}\left(I(\phi_{1})-I(\phi_{2})\right)e^{-q+p}}{\left(e^{q}+e^{-q}
+\eta\right)\left(e^{p}\int^{x}I(\phi_{1})e^{-2\phi_{1}}dy+e^{-p}\int^{x}I(\phi_{2})e^{2\phi_{2}}dy\right)}, \label{l3a}
\end{equation}

and
\begin{equation}
e^{-\Lambda}=\frac{1}{8\s}\frac{(\pa_{t_{-3}}q) e^{-p}-4\s^{2}(\pa_{x}q)\int^{x}I(\phi_{1})e^{-2\phi_{1}}dy
-2\s^{2}\left(I(\phi_{1})-I(\phi_{2})\right)e^{-q-p}}{\left(e^{p}\int^{x}I(\phi_{1})e^{-2\phi_{1}}dy+e^{-p}\int^{x}I(\phi_{2})e^{2\phi_{2}}dy\right)}. \label{l3b}
\end{equation}

 \item $t_N=t_{-5}$
 
 Similarly  we have worked out the type II Backlund transformation for $t=t_{-5}$ evolution equation (\ref{tm5}).
 The general construction  (\ref{7})  leads to 
 \br
 A_{t_{-5}} &=& e^{-2\phi} E_{\a}^{(-3)} + e^{2\phi}E_{-\a}^{(-2)} -2 I(\phi) h^{(-2)} -4 e^{-2\phi} \int^x I(\phi ) e^{2\phi} E_{\a}^{(-2)}\nonu \\
 &+& 4 e^{2\phi } \int^x I(\phi) e^{-2\phi} E_{-\a}^{( -1)} -4 W(\phi ) h^{(-1)} -8 e^{-2\phi } \int^x W(\phi ) e^{2\phi} E_{\a}^{(-1)} \nonu \\
 &+& 8e^{2\phi} \int ^x W(\phi) e^{-2\phi } E_{-\a}^{(0)}
 \label{am5}\er
where 
   \br
   W(\phi)= \int^x  \( e^{2\phi} \int^y I(\phi) e^{-2\phi} + e^{-2\phi} \int^y I(\phi) e^{2\phi}  \)\label{w}
 \er
 and the gauge transformation (\ref{k})  for $t= t_{-5}$ with  $K$  given by (\ref{kmatrix}) yields  the following Backlund transformation,
 \br
 \pa_{t_{-5}}q &=& -4\s^2 (W(\phi_2) - W(\phi_1))e^{-q} - 4\s e^{\L}(e^q + e^{-q} + \eta) \int^x W(\phi_1 ) e^{-2\phi_1} \nonu \\
 &+& 16 \s e^{-\L} \int ^x W(\phi_2) e^{2\phi_2}
 \label{t55}
 \er
 \br
 \pa_{t_{-5}}(p- \L) &=& -4 \s e^{\L+q} \int^x W(\phi_1) e^{-2\phi_1} + 4 \s e^{\L -q} \int^x W(\phi_2 ) e^{-2\phi_2}, \nonu \\
 \label{l55}
 \er
 \br 
 \pa_{t_{-5}}q &=& -4\s^2 (W(\phi_2) - W(\phi_1))e^{q} - 4\s e^{\L}(e^q + e^{-q} + \eta) \int^x W(\phi_2 ) e^{-2\phi_2} \nonu \\
 &+& 16 \s e^{-\L} \int ^x W(\phi_1) e^{2\phi_1}
 \label{t55b}
 \er
 \br 
 \pa_{t_{-5}}(\L-p)(e^{q}+e^{-q}+\eta)&=&-\pa_{t_{-5}}q(e^{q}-e^{-q})-16\s e^{-\L-q}\int^{x}W(\phi_{1})e^{2\phi_{1}}\nonu\\& &+16\s e^{-\L+q}\int^{x}W(\phi_{2})e^{2\phi_{2}} . 
 \label{l55b}
 \er
 By direct calculation we have verified that the compatibility of equations (\ref{txa}), (\ref{txb}) with (\ref{t55}), (\ref{l55}), \eqref{t55b}, \eqref{l55b} indeed 
 leads to the equations of motion (\ref{tm5}).  It thus follows that
 \begin{equation}
e^{\Lambda}=-{\frac{1}{4\s}} {{{\pa_{t_{-5}}}q e^p + 8 \s^2 \pa_x q \int^x W(\phi_2) e^{2\phi_2} + 4 \s^2 (W(\phi_2) - W(\phi_1))e^{2\phi_2}  }
\over { (e^q+ e^{-q} + \eta ) ( e^p \int^x W(\phi_1) e^{-2\phi_1} + e^{-p} \int^x W(\phi_2 ) e^{2\phi_2} ) }}, \label{l5a}
\end{equation}
and
\begin{equation}
e^{-\Lambda}=\frac{1}{16\s}  {{{\pa_{t_{-5}}}q e^{-p} - 8 \s^2 \pa_x q \int^x W(\phi_1) e^{-2\phi_1} + 4 \s^2 (W(\phi_2) - W(\phi_1))e^{-2\phi_1}  }
\over {  ( e^p \int^x W(\phi_1) e^{-2\phi_1} + e^{-p} \int^x W(\phi_2 ) e^{2\phi_2} ) }}\label{l5b}
\end{equation}

 \end{itemize}

 \section{ Solutions}
 We  now discuss  some  solutions  for  the Type-II Backlund transformation  already derived in the previous section.  
 \subsection{Vacuum-Vacuum solution}
 Let us consider
 \br 
 \phi_{1}=0 \hspace{1.5cm} \phi_{2}=0.
 \er
 It thus follows that, for  $N=-5,-3,-1,3,5$
 \br 
 e^{2\L}=-\frac{4}{2+\eta}. \label{lamb}
 \er
 Writing  $\eta=-(e^{2\t}+e^{-2\t})$  eqn. (\ref{lamb})  becomes
 \br 
 e^{2\L}=\frac{1}{\sinh^{2} \t}.
 \er
 The vacuum-vacuum solution therefore leads to a constant  $\L$ \cite{cor-zamb2}. 
 \subsection{Vacuum to one-soliton Solution}
 Let us consider the following field configurations {\footnote{ Solutions $\phi_1$ and $\phi_2$  with appropriate $\rho(x, t_N)$ solves all equations within the mKdV hierarchy, i.e., eqns.  (\ref{t3})-(\ref{t7}) and   (2.17)-(\ref{tm5}) , etc  for $N=3,5,7, \cdots$ and $-3,-5, \cdots$, 
}}
  \br
 \phi_1 = 0, \qquad \phi_2 = \ln \( {{1+\rho }\over {1-\rho}} \)
 \label{sh1sol}
 \er
where  $ \rho(x, t_N) = \exp \( 2kx +2( k)^{N}t_{N} \) $.

 It therefore follows for the sinh-Gordon model where $t_N = t_{-1}$  that inserting (\ref{sh1sol}) in (\ref{l1}), 
 \br
 e^{\L} &=& {{2}\over {\s}} {{\pa_{t_{-1}} \phi_2e^{\phi_2} -\s^2\pa_x \phi_2}\over { (e^{\phi_2} - e^{-\phi_2})( e^{\phi_2} + e^{-\phi_2}+ \eta) }} = 
 2(1+\rho){{(1+\rho - k^2 \s ^2(1-\rho))}\over {k \s ((2+\eta ) + \rho^2 (2 - \eta ))}}, \nonu \\ 
 \\
  e^{-\L} &=& -{{1}\over {2\s}} {{\pa_{t_{-1}} \phi_2 e^{-\phi_2}-\s^2\pa_x \phi_2}\over { (e^{\phi_2} - e^{-\phi_2}) }} = 
  {{(-1 + k^2 \s Â²) + \rho ( 1+ k^2 \s Â²)}\over {2 k \s (1 + \rho )}}
 \er
 The identity 
 \br
  e^{\L} e^{-\L} -1 = {{ (1 + \eta k^2 \s ^2 + k^4 \s ^4)}\over { 2+\eta  + \rho^2 (2-\eta )}} \( {{1 - \rho^2}\over {k^2 \s^2}} \)= 0
  \label{id1}
  \er
  implies
  \br
  1 + \eta k^2 \s ^2 + k^4 \s ^4 = 0 \label{id11}
  \er 
  which
 leads  to values  of the Backlund parameter $\s$.  Parametrizing $\s = e^{\varphi},\quad k = e^{\theta}, \quad  \eta = -(e^{2\tau } + e^{-2\tau })$ 
 we find four solutions for the identity (\ref{id11})
 \br
 \varphi & =& -\theta \pm  \tau \label{sol1} \\
 \varphi & =& -\theta \pm   \tau  +i\pi \label{sol2} \\
  \er
 We have also verified, using Mathematica  program, that the {\it very same condition}  (\ref{id11}) is required  for $t=t_{-3}$ and $t=t_{-5}$ 
 obtained from (\ref{l3a}), (\ref{l3b}) and  from (\ref{l5a}), (\ref{l5b}) respectively.

 For the mKdV equation where $t_N = t_3$, we find after substituting (\ref{sh1sol})  into (\ref{l1a}) and (\ref{l1b}),
 \br
  e^{\L} e^{-\L} -1 = {{{( 1 + \eta k^2 \s ^2 + k^4 \s ^4) }\over {2+\eta +(2-\eta) \rho^2}}}  \( A_3 + B_3\rho^2 \)  =0 \label{id2}
\er
where
\br
A_3 = 2+\eta + k^2 \s ^2, \qquad B_3 =  -2 + \eta + k^2 \s ^2.
\er
  Similarly for $t_N = t_5$,
 \br
  e^{\L} e^{-\L} -1 = {{{( 1 + \eta k^2 \s ^2 + k^4 \s ^4) \over{ 2+\eta +(2-\eta) \rho^2 }}}} \( {{A_5+B_5\rho^2}\over {\eta^2}}\)=0 \label{id3}
\er
where
\br
A_5&=& -\eta^3 + \eta^2(k^2\s^2 -2) +\eta k^2\s^2(2+ k^2\s^2)-k^2\s^2(1+ k^4\s^4), \nonu \\
B_5&=&  \eta^3 - \eta^2(k^2\s^2 +2) +\eta k^2\s^2(2- k^2\s^2)+k^2\s^2(1+ k^4\s^4)
\er
 Identities  (\ref{id2}) and (\ref{id3})  are both satisfied  by condition (\ref{id11})  and henceforth  the same result 
 (\ref{sol1}) and (\ref{sol2}) holds for all cases considered, i.e., $N=-5,-3,-1,3,5$. It is interesting to observe that in all cases considered the   
 result for the auxiliary field $\L$ is precisely the same, namely
 \br
 e^\L = 2e^{\tau} {{{1+\rho }\over {\pm 1\mp e^{2\tau} + (1+ e^{2\tau})\rho}}}, \quad {\rm for } \quad \varphi = -\theta \mp \tau
 \er
 or 
 \br
 e^\L = -2e^{\tau} {{{1+\rho }\over {\pm 1\mp e^{2\tau} + (1+ e^{2\tau})\rho}}}, \quad {\rm for } \quad \varphi = -\theta \mp \tau + i\pi
 \er
 Most probably this is a direct consequence of the universality of the Lax  space component within the hierarchy.

 \subsection{Scattering of one-soliton Solutions } Let  us consider the   one-soliton  solutions  of the form
 \br
 \phi_1 = \ln \( {{1+\rho(x, t_N)}\over {1-\rho(x, t_N)}}\), \qquad \phi_2 = \ln \( {{1+R \rho(x, t_N)}\over {1-R\rho(x, t_N)}}\)
 \label{sol}
 \er
 where  $\rho(x, t_N) = \exp \(2kx +2( k)^Nt_N \)$.  It thus follows that
 \br
 \pa_{t_N} q = \pa_{t_{N}} (\phi_1-\phi_2) = 4k^N \rho \( {{1}\over {1-\rho^2}} -  {{R}\over {1-R^2\rho^2}}\)
 \label{deriv}
 \er
 and similarly for $\pa_x q$.
  
 For the sinh-Gordon, substituting $\phi_1$ and $\phi_2$ into the equation (\ref{l1})  we find  that 
  \br
  & &  \(1-  e^{\L} e^{-\L} \){{ {k^2\s^2(1+R)^2 } \over {(1-\rho^2)(1-R^2 \rho^2)}}}   \nonu \\
 &= & {{{ (1+k^2\eta \s^2 + k^4\s^4)(1+R^2)   -2R(1 - k^2(4+\eta)\s^2 +
 k^4\s^4)}\over { 2+\eta -(-2+8R+R^2(-2+\eta)+\eta )\rho^2 + (2+\eta)R^2\rho^4}}} =0 \nonu \\
 \label{e}
 \er
 and henceforth the  delay $ R$ satisfies
 \br
  (1+k^2\eta \s^2 + k^4\s^4)(1+R^2)   -2R(1 - k^2(4+\eta)\s^2 +
 k^4\s^4) = 0. \label{R}
 \er
 The two solutions are 
\br
 R_{1}=\frac{1-4k^{2}\s^{2}-k^{2}\eta\s^{2}+k^{4}\s^{4}-2k\s(k^{2}\s^{2}-1)\sqrt{-2-\eta}}{1+k^{2}\eta\s^{2}+k^{4}\s^{4}},
\er
and
\br
R_{2}=\frac{1-4k^{2}\s^{2}-k^{2}\eta\s^{2}+k^{4}\s^{4}+2k\s(k^{2}\s^{2}-1)\sqrt{-2-\eta}}{1+k^{2}\eta\s^{2}+k^{4}\s^{4}},
\er
  We can rewrite $R_{1}$ and $R_{2}$ in a more simple form if we make the following parametrization
\begin{equation}
\eta=-e^{2\tau}-e^{-2\tau}, \qquad 
k=e^\theta,\qquad 
\s=e^{\varphi}.
\end{equation}
With these choices we find that
\begin{equation}
R_{1}=\frac{e^{\tau}-e^{\theta+\varphi}+e^{\theta+2\tau+\varphi}-e^{2\theta+\tau+2\varphi}}{e^{\tau}+e^{\theta+\varphi}-e^{\theta+2\tau+\varphi}-e^{2\theta+\tau+2\varphi}}=-\coth\left(\frac{-\theta-\varphi-\tau}{2}\right)\tanh\left(\frac{\theta+\varphi-\tau}{2}\right),
\end{equation}
and
\begin{equation}
R_{2}=\frac{1}{R_{1}}.
\end{equation}
  Similarly from  (\ref{l3a}), (\ref{l3b}) for $t=t_{-3}$  and  (\ref{l5a}), (\ref{l5b}) for $t=t_{-5}$  we find  the same result (\ref{e}) and the same 
  two solutions for $R$.  For the mKdV the condition (\ref{R}) remains  true (see appendix A) and leads to the same two solutions for $R$.
We therefore conclude that the delay  $R$ is the same for the $t= t_{-5}, t_{-3}, t_{-1} $ and $t_3$.   Such result might as well  be true for all times.

 \section{Conclusions and Further Developments}
 
 We have verified, within the mKdV hierarchy  that there exist a single gauge transformation that generates the Backlund transformation for the first
 few integrable equations  associated to both, positive and negative graded time evolution equations.   
 In particular, we have shown that  the very same gauge transformation generating  the Backlund transformation  for the sinh-Gordon model  
was verified to  generate  Backlund transformation  for  the mKdV and other higher (lower)  graded evolution equations.

 Moreover we have verified that the vacuum-vacuum, vacuum-one-soliton transitions and the scattering of one-soliton
 solutions of the Backlund transformation are correlated  according to the different time evolution parameters.  
 In fact,  the Backlund parameters  are fixed (for a specific model) 
 and remain  the same within all equations of motion  considered (i.e. are independent of the model in question).

 We conclude  this paper by establishing  the conjecture that the type I as well as  type II  Backlund transformation
for a given hierarchy may be systematically derived  from a fundamental principle of the universality of the  space Lax operator.
This fact  induces the universality of    a gauge transformation which maps  the
 two field configurations related by the Backlund transformation for the entire hierarchy.

More general hierarchies beyond  the mKdV hierarchy  may be considered also.  
 In particular the type II Backlund transformation for the $N=1$ super  sinh-Gordon  models was recently derived in \cite{alexis} and may be 
 extended to other  equations of motion within the hierarchy.    For the non linear Schroedinger equation (NLS), the  Backlund transformation   
 was derived  in terms of a Ricatti equation  \cite{konno}.  It would be interesting to verify our conjecture within the NLS hierarchy. 
 
 \vskip .4cm \noindent
{\bf Acknowledgements} \\
We  would like to thank  Capes, CNPq and Fapesp for support. We also thank N.I.Spano for discussions.

\section{Appendix A -- One Soliton scattering for mKdV $t=t_3$ }

For the mKdV equation $t=t_3$ the  one soliton scattering  with solitions (\ref{sol})  yields 
\br 
 & &e^{\L}e^{-\L}-1=  \nonu \\
& & \frac{ \( (1+R^{2})(1+\eta k^{2}\s^{2}+k^{4}\s^{4})-2R(1-(4+\eta)k^{2}\s^{2}+k^{4}\s^{4})\)}{d_{1}d_{2}} \sum_{n=0}^{6}c_{n}\rh(x,t)^{2n} =0\nonu
\er
with

\br
 c_0 &=& -(2+\eta+k^{2}\s^{2}) \nonu \\
 c_1 &=& (2+3\eta+3k^{2}\s^{2})+(12+2\eta-2k^{2}\s^{2})R
+(-2+\eta+k^{2}\s^{2})R^{2} \nonu \\
 c_2 &=& (2-3\eta-3k^{2}\s^{2})+(-20-6\eta+6k^{2}\s^{2})R-4(4+\eta+k^{2}\s^{2})R^{2}\nonu \\
 &+&(4-2\eta+2k^{2}\s^{2})R^{3}\nonu\\
 c_3 & =& (-2+\eta+k^{2}\s^{2})+(4+6\eta-6k^{2}\s^{2})R+6(6+\eta+k^{2}\s^{2})R^{2}+\nonu\\
          &+&(4+6\eta-6k^{2}\s^{2})R^{3}+(-2+\eta+k^{2}\s^{2})R^{4}\nonu\\
c_4 &=& (2-3\eta-3k^{2}\s^{2})-2(-2+\eta-k^{2}\s^{2})R-4(4+\eta+k^{2}\s^{2})R^{2}\nonu \\
&+&(-20-6\eta+6k^{2}\s^{2})R^{3}\nonu\\
 c_5 &=& (-2+\eta+k^{2}\s^{2})R^{2}+(12+2\eta-2k^{2}\s^{2})R^{3}+(2+3\eta+3k^{2}\s^{2})R^{4}\nonu\\
 c_6 & =& -(2+\eta+k^{2}\s^{2})R^{4}\nonu
\er
\noindent
and
\br
 d_{1}&=&2+\eta -(-2+8R+R^2(-2+\eta)+\eta )\rho^2 + (2+\eta)R^2\rho^4\nonu\\
 d_{2}&=&[1-R-2k^{2}\s^{2}+\(-1-2k^{2}\s^{2}+(1+2k^{2}\s^{2})R^{2}\)\rho^{2}\nonu \\
 &+&(R-(1-2k^{2}\s^{2})R^{2})]^{2}\nonu
\er
and the condition for the delay coincide with (\ref{R}), i.e.,
\br
 (1+R^{2})(1+\eta k^{2}\s^{2}+k^{4}\s^{4})-2R(1-(4+\eta)k^{2}\s^{2}+k^{4}\s^{4}) =0
\er

\section{Appendix B -- Obtaining the K matrix}

In order to find the Backlund transformations to the equations of mKdV hierarchy, we suppose that the $K$ matrix is given by
\br 
K= \b + {{{\g}\over {\l}}} = \begin{bmatrix} \b_{11}+\frac{\g_{11}}{\l} & \b_{12}+\frac{\g_{12}}{\l} \\  \b_{21}+\frac{\g_{21}}{\l} &  \b_{22}+\frac{\g_{22}}{\l}\end{bmatrix},
\er
and we can write
\br
\pa_{\mu}K=KA_{\mu}(\phi_{1})-A_{\mu}(\phi_{2})K,
\label{gt}
\er
where $x_{\mu}=x,t_{N}$.\

Then, we start by the spacial part because $A_{x}(\phi)$ is the same for all equations of the hierarchy, having the following form
\br 
A_{x}(\phi)=\begin{bmatrix} \pa_{x}\phi & 1 \\ \l & -\pa_{x}\phi  \end{bmatrix}.
\er
Calculating the terms of equation \eqref{gt}, with $x_{\mu}=x$, we obtain a matrix whose 11 element is given by
\br 
& & \l^{-1}: \hspace{0.2cm} \pa_{x}\g_{11}=\g_{11}\pa_{x}q-\g_{21}\\  
& & \l^{0}\hspace{0.2cm}: \hspace{0.3cm} \pa_{x}\b_{11}=\b_{11}\pa_{x}q+\g_{12}-\b_{21}\\ 
& & \l^{1}\hspace{0.2cm}: \hspace{0.3cm} \b_{12}=0;
\er
the 12 matrix element in turn is
\br 
& & \l^{-1}: \hspace{0.2cm} \pa_{x}\g_{12}=-\g_{12}\pa_{x}p+\g_{11}-\g_{22} \label{1.7} \\
& & \l^{0}\hspace{0.2cm}: \hspace{0.2cm} \pa_{x}\b_{12}=-\b_{12}\pa_{x}p+\b_{11}-\b_{22}.
\er
We also have the 21 matrix element
\br 
& & \l^{-1}: \hspace{0.2cm} \pa_{x}\g_{21} =\g_{21}\pa_{x}p\\
& & \l^{0} \hspace{0.2cm}: \hspace{0.3cm} \pa_{x}\b_{21}=\b_{21}\pa_{x}p+\g_{22}-\g_{11} \label{1.10}\\
& & \l^{1} \hspace{0.2cm}: \hspace{0.3cm} \b_{22}=\b_{11};
\er
and the 22 matrix element given by
\br 
& & \l^{-1}: \hspace{0.2cm} \pa_{x}\g_{22}=-\g_{22}\pa_{x}q+\g_{21}\\  
& & \l^{0}\hspace{0.2cm}: \hspace{0.3cm} \pa_{x}\b_{22}=-\b_{22}\pa_{x}q+\b_{21}-\g_{12}\\ 
& & \l^{1}\hspace{0.2cm}: \hspace{0.3cm} \b_{12}=0.
\er
 Using the equations (9.6), (9.11) together with equations (9.4) and (9.13) we obtain 
\br 
& & \b_{12}=0;\\
& & \b_{22}=\b_{11}=\text{constant}\equiv b_{11};\\
& & \b_{21}=b_{11}\pa_{x}q+\g_{12}. 
\er

Let us now, consider the time component of equation \eqref{gt}. Putting  
\br 
A_{t_{-1}}=\begin{bmatrix}0 & \frac{1}{\l}e^{-2\phi}\\ e^{2\phi} & 0 \end{bmatrix}.
\er
in the equation \eqref{gt} we find for the $\l^{-2}$  coefficient of the 11  matrix  element  to be 
\br 
  \l^{-2}: \hspace{0.2cm} \g_{21}=0
\er

The same result,  $\g_{21}=0$ follows for $A_{t_{-3}}$ and $A_{t_{-5}}$.

Taking now  $A_{t_{3}}$ as
\br 
A_{t_{3}}(\phi)=\begin{bmatrix}
v \l+\frac{1}{4}\pa^{2}_{x}v -\frac{1}{2}v^{3} & \l-\frac{1}{2}v^{2}+\frac{1}{2}\pa_{x}v \\
\l^{2}-\frac{1}{2}v^{2}\l-\frac{1}{2}\pa_{x}v\l & -v \l-\frac{1}{4}\pa^{2}_{x}v +\frac{1}{2}v^{3}\\
\end{bmatrix}
\er
in equation \eqref{gt} we obtain that the term proportional to $\l^{-1}$ of the matrix element 21 to be
\br 
\pa_{t_{3}}\g_{21}=(\frac{1}{4}\pa^{2}_{x}v_{1} +\frac{1}{4}\pa^{2}_{x}v_{2}-\frac{1}{2}v_{1}^{3}-\frac{1}{2}v_{2}^{3}) \g_{21},
\er
which can be rewritten as  
\br 
\pa_{t_{3}}\g_{21}=\pa_{t_{3}}(\phi_{1}+\phi_{2})\g_{21} \label{xx}
\er
Again $\g_{21}=0$  is  a particular solution.  The same type of equation (\ref{xx}) for $\g_{21} $  follows for 
 $N=5$.
In this work, we will assume 
\br 
\g_{21}=0,
\er
but the more general assumption still to be   studied.

With this choice we conclude using the equation (9.4) and (9.12) that
\br 
\g_{11}=c_{11}e^{q}, \qquad \g_{22}=c_{22}e^{-q}.
\label{g11}
\er
We have then two possible solutions:
\begin{itemize}
\item $\g_{11}=\g_{22}$;
\end{itemize}
This implies that  $c_{11}= c_{22}=0$  and henceforth from (\ref{1.7}) and (\ref{1.10}), 
\br 
 \b_{21}=b_{21}e^{p}  \qquad 
 \g_{12}=c_{12}e^{-p} \label{9.25}
\er
Setting  $b_{11}\equiv 1$, $b_{21}=c_{12}=-\frac{\b}{2}$ we obtain
\br 
K=\begin{bmatrix}
1 & -\frac{\b}{2\l}e^{-p}\\
-\frac{\b}{2}e^{p} & 1
\end{bmatrix}
\er
 the K matrix that generates the {\it Type-I Backlund transformations}.
\begin{itemize}
\item $\g_{11}\neq\g_{22}$
\end{itemize}
We start by the following expression
\br 
\pa_{x}(\g_{12}\b_{21})=(\pa_{x}\g_{12})\b_{21}+\g_{12}(\pa_{x}\b_{21}) 
\er
and using the equations \eqref{1.7},\eqref{1.10} and \eqref{g11} we have
\br 
\pa_{x}(\g_{12}\b_{21})=b_{11}(c_{11}e^{q}-c_{22}e^{-q})\pa_{x}q 
\er
and assuming $c_{11}=c_{22}$, results in
\br 
\g_{12}\b_{21}=b_{11}c_{11}(e^{q}+e^{-q}+\eta). \label{9.30}
\er
with $\eta $is  an integration constant. Generalizing \eqref{9.25}
\br 
\b_{21}=b_{21}e^{p-\L}
\er
 by    \eqref{1.10} gives
 \br
 \pa_x \L = {{c_{11}}\over {b_{21}}}e^{\L-p}(e^q -e^{-q})
 \er
 which has the form  \eqref{type-iixb}.  Expression \eqref{9.30} implies:
\br 
\g_{12}=\frac{{b_{11}c_{11}}{b_{21}}}e^{\L-p}(e^{q}+e^{-q}+\eta).
\er

Putting $b_{11}\equiv 1$, $c_{11}=-\frac{1}{\s^{2}}$ and $b_{21}=-\frac{2}{\s}$, we obtain
\br 
K=\begin{bmatrix}
1-\frac{1}{\s^{2}\l}e^{q} & \frac{1}{2\s\l}e^{\L-p}(e^{q}+e^{-q}+\eta) \\
-\frac{2}{\s}e^{p-\L} & 1-\frac{1}{\s^{2}\l}e^{-q} 
\end{bmatrix}
\er
which is the $K$ matrix that generates the {\it Type-II Backlund transformations} .

As a  conclusion of this apppendix, we have shown that for $t= t_{-1}, t_{-3} $ and $t_{-5}$  
we have  universality of the matrix $K$ while for $t=t_3, t_5, \cdots $ we 
have it as a particular solution.



\end{document}